\documentclass[aps,prb,twocolumn,superscriptaddress]{revtex4-2}

\usepackage{amsmath}
\usepackage{amssymb}
\usepackage{graphicx}
\usepackage{bm,hyperref,bbold}
\usepackage{xcolor}

\definecolor{DarkGreen}{RGB}{1,50,32}

\def\be{\begin{equation}}

\def\ba{\begin{eqnarray}}
\def\ea{\end{eqnarray}}

\def\pp{{\bm p }}

\begin{document}
\title{Magnetic vortex control with current-induced axial magnetization in centrosymmetric Weyl materials}

\date{\today}
\author{J. G. Yang}
\affiliation {Department of Physics, University of Virginia, Charlottesville, Virginia 22904, USA}
\author {Yaroslav Tserkovnyak}
\affiliation{Department of Physics and Astronomy and Bhaumik Institute for Theoretical Physics, University of California, Los Angeles, California 90095, USA}
\author{D. A. Pesin}
\affiliation {Department of Physics, University of Virginia, Charlottesville, Virginia 22904, USA}
\begin{abstract}
We consider magnetic Weyl metals as a platform to achieve current control of magnetization textures with transport currents, utilizing their underlying band geometry. We show that the transport current in a Weyl semimetal produces an axial magnetization due to orbital magnetic moments of the Weyl electrons. The associated axial magnetization can generate a torque acting on the localized magnetic moments. For the case of a magnetic vortex in a nanodisk of Weyl materials, this current-induced torque can be used to reverse its circulation and polarity. We discuss the axial magnetization torques in Weyl metals on general symmetry grounds, and compare their strength to current-induced torques in more conventional materials.
\end{abstract}
\maketitle 

\textit{Introduction.---} Discrete degrees of freedom in condensed matter systems have firmly established themselves as perpetual candidates for information storage units. Their microscopic versions - single spins, single charges in single-electron boxes - have a long history of being considered for qubit realizations. Discrete macroscopic degrees of freedom, predominantly those associated with magnetization and its direction, have been not merely candidates, but also workhorses of information storage, albeit classical one, see reviews ~\cite{parkin2008magneticmemory,tatara2008walldynamics}. Proposals to use mesoscopic magnetization textures in nanoscale samples as platforms for quantum information storage and manipulation have also emerged~\cite{braun1994blochwall,takagi1996quantum,caretta2018skyrmions,psaroudaki2021skyrmionqubits,heyderman2021mesoscopic,ezawa2022qubits,silva2023domainwalls}. 

Associated with these proposals is the question of control of small-scale magnetization textures. Accomplishing this control with electric currents promises practical benefits coming from scalability of the architectures, and reduced power consumption, see Ref.~\cite{fert2023review} for a recent review of the field. In the realm of spintronics of conventional materials~\cite{vzutic2004spintronics}, as opposite to topological ones, there is a number of well known ways to approach magnetization control with current. The list includes spin-transfer torques in spin-valve-type devices~\cite{ralph2008stt}, spin current injection via spin Hall effect~\cite{sinova2015she}, and current-induced torques in noncentrosymmetric systems with strong spin-orbit coupling~\cite{manchon2019currentinduced}.  

In this work we consider magnetic Weyl metals as a possible material candidate for realization of information storage and manipulation in nanoscale systems. We show that there is a new source of a current-induced spin torques in these materials related to the current-induced axial magnetization, which changes sign between the valleys of the Weyl material. The axial current associated with the axial magnetization induces nonequilibrium spin polarization of itinerant carriers. This spin polarization is capable of controlling textures of the underlying magnetization of localized spins, which is responsible for the equilibrium magnetism in the sample. This mechanism of texture control is distinct from the existing proposals of current-induced spin torques in Weyl materials due to the axial Hall current produced by the pseudomagnetic field of magnetic textures~\cite{kurebayashi2019theory}, or torques stemming from the chiral anomaly~\cite{hannukainen2021electric}. All three mechanisms are compared in the Discussion presented at the end of this paper. 

As a specific application of the developed theory we consider magnetic vortex control in thin magnetically soft nanodisks, see Ref.~\cite{guslienko2008review} for a review of the subject. The practical motivation behind this choice of a texture comes from the fact that the vortex in a nanodisk is a compact object with discrete states determined by its core polarization and chirality. The vortex state develops to minimize the magnetic dipolar energy in nanodisks of size roughly exceeding their magnetic exchange length. This suggests that nanodisks assembled into an array will have weak interaction due to their stray fields, which is beneficial for high density information storage. Below we will show that the torques due to the current-induced axial magnetization in Weyl metals can efficiently flip vortex chirality, and even its polarization under the right circumstances.

\textit{Electromagnetic fields and pseudofields in a magnetic Weyl metal.---} We view a magnetic Weyl metal in the spirit of the $s-d$ exchange model, which includes a subsystem of localized electrons responsible for the magnetization, and a system of itinerant electrons carrying transport currents. Our goal is to find a way to control the magnetization of localized electrons with the transport currents. 

We use the prototypical model of a magnetic Weyl metal with only two Weyl points with opposite chiralities close to the Fermi level. Such a model preserves the inversion symmetry, but the time-reversal symmetry is broken by the magnetization, $\bm M$, of the localized electrons. 

The Hamiltonian of the model is given by \cite{armitage2018weyl,kurebayashi2019theory}
\begin{equation}\label{eq:Weylatom}
   H_{\rm{w}}= \int d^{3}r \psi ^{\dagger}(\bm{r})\left[  v\tau_{z}\bm \sigma \cdot \bm{p} -J\tau_0 \bm \sigma\cdot \bm m\right] \psi(\bm{r}). 
\end{equation}
where $\psi (\bm r)$ are the field operators for electrons, $ \bm \sigma$ is a vector of Pauli matrices acting in the space spanned by the Weyl bands, which we will take to coincide with the actual spin, while $\tau_z$ and $\tau_0$ act in the valley space. The unit matrix $\tau_0$ will not be explicitly written from here on. Furthermore, in Eq.~\eqref{eq:Weylatom} $ v$ is the Fermi speed, $J$ is the exchange energy constant between itinerant electrons and localized spins, $\bm{m}\equiv \bm M/M_s$ is a unit vector in the direction of the localized  magnetization. For definiteness, we will assume $v>0$, and denote the $\tau_z=\pm$ valleys with the chirality index $\chi=\pm$. 

As simple as it is, the model~\eqref{eq:Weylatom} might pertain to the case of EuCd$_2$As$_2$, either in a small external magnetic field~\cite{soh2019eucdas}, or grown in the ferromagnetic phase, as well as $\mathrm{K}_2\mathrm{Mn}_3(\mathrm{As}\mathrm{O}_4)_3$\cite{nie2022idealweyl}. But one should keep in mind recent evidence that EuCd$_2$As$_2$ is in fact a narrow-gap semiconductor~\cite{akrap2023eucdas}.

In this work, we will consider a Weyl magnet in which there exist both a static magnetic texture, $\bm m=\bm m (\bm r)$, and transport current density, $\bm j_{\rm{tr}}(\bm r)$. Our aim is to find a way to manipulate the texture with the transport current. A general way to achieve this goal follows from Eq.~\eqref{eq:Weylatom}, which shows that the magnetization of the localized electrons couples to the spin polarization of the itinerant ones, which induces an effective Zeeman field 
\begin{align}\label{eq:effectivefield}
    \bm B_{\rm{eff}}=\frac{J}{M_s} \langle\psi^\dagger \bm \sigma \psi\rangle,
\end{align}
where the $\langle\ldots\rangle$ denotes the average with respect of the density matrix of the itinerant electrons, and $M_s$ is the sturation magnetization of the localized electrons. In turn, the spin polarization of the itinerant electrons is identical to the axial current, $\bm j_5$, defined as the difference in the individual valley currents:
\begin{align}\label{eq:spinfromaxialcurrent}
  \langle\psi^\dagger \bm \sigma \psi\rangle=\frac{1}{ev}(\bm j_{+}-\bm j_{-})\equiv \frac{1}{ev}\bm j_5, 
\end{align}
where $\bm j_{\chi}$ is the current in the valley with chirality $\chi$, and $e<0$ is the charge of the electron. The conclusion is that one must search for valley-asymmetric currents to control magnetization, or magnetic textures. 

In the presence of a nonuniform magnetization, one has to take into account electric fields, magnetic fields coming from the magnetization and the transport current, as well as pseudomagnetic fields from magnetization gradients while considering a Weyl metal. The pseudomagnetic field appears because in Hamiltonian~\eqref{eq:Weylatom} magnetization couples to the electrons in the two valleys as an axial vector potential, having the opposite signs in the opposite valleys, $e\bm{A}_{5}=\frac{J}{v}\bm m$. 
Then it is clear that in the presence of a spatially varying magnetization the axial vector potential $\bm A_5$ can develop a non-zero curl, and the corresponding pseudomagnetic field is
\begin{align}\label{eq:B5}
    e\bm B_5=\frac{J}{v}\bm \nabla\times \bm m.
\end{align}

There are many physical effects brought about by the fields mentioned above. A review of the pseudofield physics in Weyl metals can be found in Ref.~\onlinecite{araki2020review}. Fortunately, not all of them are equally important in the present context, and we would like to discuss qualitatively which parts of physics need to be included into the qualitative theory, before we actually attempt it. We will focus on the phenomena specific to Weyl magnets, leaving aside phenomena associated with the usual diffusive transport in metals. 

First of all, there is a number of known phenomena that can be used to control the magnetization. The two most famous examples are the current-induced magnetization in noncentrosymmetric samples, and the spin Hall effect. In the present work we consider centrosymmetric crystals, such that the current-induced spin polarization does not appear, and assume that the spin-Hall effect does not exist, which is true for the model of Eq.~\eqref{eq:Weylatom}.

It has already been noticed in Ref.~\cite{kurebayashi2019theory} that an axial magnetic field drives an axial Hall current in the presence of a transport electric field, which leads to a contribution to a net spin polarization. In this work we will consider transport currents flowing along the axial magnetic field, hence the axial Hall current can be neglected. 

Furthermore, in the presence of a transport electric field, the pseudomagnetic field can drive an anomaly-type term in the equation for the local (number) density of electrons~\cite{qi2013anomaly}, 
\begin{align}\label{eq:B5anomaly}
\partial_t n=\frac{e^2}{2\pi^2\hbar^2} \bm E\cdot \bm B_5. 
\end{align}
Physically, this term stems from the divergence of the space-dependent current driven by intrinsic nonuniform Hall conductivity proportional to the separation of the Weyl nodes in the momentum space~\cite{wan2012weyl,burkov2011weyl,ran2011weylahe}, made space-dependent by the space-dependent magnetization. The change in the electronic density implied by Eq.~\eqref{eq:B5anomaly} is essentially forbidden in metallic samples due to screening. Perturbations that violate local charge neutrality are effectively relaxed in three-dimensional metals on the scale of Maxwell relaxation time, determined by the inverse Drude conductivity: $\tau_M\sim\epsilon_0/\sigma_D$. Even for a reasonably low conductivity of $\sigma_D\sim 10^{6}\,\mathrm{\Omega}^{-1}\mathrm{m}^{-1}$~\cite{roychowdhury2023eucd2as2}, this relaxation time is of order of $10^{-17}\,\mathrm{s}$, hence such perturbations can be completely disregarded. 

Finally, a discussion of a Weyl material is incomplete without mentioning the effect of surface Fermi arcs. In a magnetic nanodisk type of a sample considered below, the particular shape and length of a Fermi arc is determined by the projection of the magnetization on a sample surface in the real space. This implies that the energy of the surface electronic subsystem depends on magnetization orientation. It represents a type of surface anisotropy that describe a tendency to orient magnetization perpendicular to the surface. This tendency is at odds with the effect of dipolar interactions, whose energy is minimized when the magnetization is oriented along the surface, and no `magnetic charges' are produced. 

To roughly determine whether the surface states need to be taken into account in the energy balance, one can compare the energy of the system for magnetization along the surface, when the dipolar energy is minimized, but the Fermi arc energy is maximum, and the energy when the magnetization is perpendicular to the surface, in which case the Fermi arc energy is minimal, while the dipolar energy is the largest. Obviously, the exact balance depends on the shape of the sample, so we only aim to estimate for sample of what size the Fermi arcs are important. We note that the energy associated with a Fermi arc depends on the corresponding surface state spectrum, and on the occupation of the surface states. The surface state spectrum, determining the shape of a Fermi arc, can be quite involved, with spiraling around a Weyl point projection on the surface Brillouin zone~\cite{andreev2015arcs}, and depends on the details of the confining potential. However, the overall length of the arc in a simple model with two nodes goes linearly with the magnetization projection on the surface in real space. Further, we can assume that the band width of the Fermi arc states is of order of $J$. Then the surface energy density associated with a patch of 2D momentum space occupied by the surface states is $J^3/(2\pi\hbar v)^2$. This energy should be compared to the $\mu_0 M_s^2\ell_{ex}$, where $\ell_{ex}\sim 5\,\mathrm{nm}$ is the magnetic exchange length over which magnetization can vary near a surface. For $\mu_0 M_s\sim 1\,\mathrm{T}$, $J\sim 0.1\,\mathrm{eV}$, and $v\sim 10^5\, \mathrm{m/s}$, we see that surface state energy density is about one third of magnetic energy density, and can be ignored for our purposes. At the same time, it is obvious that the surface state energy very sensitively depends on the value of the exchange constant $J$, so one can easily encounter materials in which it has to be taken into account. While this is an interesting research direction, we do not pursue it here.

\textit{Boundary current torques in current-carrying Weyl metals.---} Given the discussion above, it is clear that our goal is to find a new source of axial current, which flows in response to a transport current. Since the axial current is a pseudovector, the linear relationship between it and the transport current (polar vector) in a centrosymmetric material is only possible either in a nonuniform situation, or near a sample boundary, where the inversion symmetry is broken by the surface. We will argue below that just the right axial currents flow as surface ``axial magnetization'' currents. 

Indeed, each valley of the band structure described by model~\eqref{eq:Weylatom} breaks effective inversion symmetry, which acts by reversing the momentum counted relative to the valley position in momentum space. The full inversion symmetry is restored when the valleys are interchanged in addition to momentum inversion. This implies that a transport current can induce magnetization in each of the valley, but the total magnetization vanishes: this situation we will refer to as having nonzero ``axial magnetization''. This means that at this level the transport current cannot affect the magnetization of the localized electrons in the centrosymmetric model we are considering. However, magnetizations in each valley, being opposite in direction, create opposite magnetization currents in regions of space where each of the magnetizations varies in space, in particular near sample boundaries. In other words, there is an axial current created by axial magnetization. This valley current is synonymous with the spin polarization of itinerant electrons, see Eq.~\eqref{eq:spinfromaxialcurrent}. Thus we expect boundary torques acting on the magnetization of the localized electrons from this mechanism. 

To describe the above mechanism of torque appearance quantitatively, we write down the expression for the magnetization in each valley as 
\begin{align}\label{eq:magnetization}
    \bm M_\chi=\int_{\bm p}\mu_{\chi,\bm p} f_{\chi,\bm p},
\end{align}
where $\int_{\bm p}\equiv\int \frac{d^3p}{(2\pi\hbar)^3}$, $f_{\chi,\pp}$ is the occupation number of a state with quasimomentum $\pp$ in valley $\chi$ and in the band (conduction or valence) that contains a Fermi surface. We do not introduce the band index explicitly not to clutter the notation. Further, $\mu_{\chi,\bm p}$ is the effective magnetic moment of an electron with quasimomentum $\bm p$. Such magnetic moment has both spin and orbital contributions, but the orbital effects are usually much stronger in Weyl materials, which is related to the fact  that the Bohr magneton contains the bare electron mass, which is very large as compared to the effective mass scale, $p_F/v$, determining the orbital magnetic moments of Weyl electrons. (See note~28 in Ref.~\cite{ma2015noa} for more details.) 

It was shown in Ref.~\cite{rou2017kineticmoments} that the orbital magnetic moments contain both an intrinsic contribution~\cite{xiao2010berry}, as well as extrinsic contributions from side jump and skew impurity scattering processes. However, for the simple isotropic model of Eq.~\eqref{eq:Weylatom} side jump and skew scattering processes vanish for isotropic impurity scattering, and only the intrinsic contribution needs to be taken into account. It would be enough to add tilt to the dispersion of the Weyl cones to get an extrinsic contribution to the magnetic moment~\cite{steiner2017ahe}.

For a single Weyl point of chirality $\chi$, we have the following expression for the intrinsic orbital angular moment~\cite{xiao2010berry}:
\begin{align}
    \bm \mu_{\chi}=\chi \frac{e\hbar v}{2p}e_{\bm p},
\end{align}
which works for both the conduction and valence bands, and where $\bm e_{\bm p}$ is the unit vector in the direction of $\bm p$. 

To calculate the axial magnetization of Weyl electrons, we use Eq.~\eqref{eq:magnetization} with the nonequilibrium distribution function of the electrons in the presence of a transport electric field, $\bm E$, $\delta f_{\bm{p}}=-\tau_{\rm{tr}} e \bm{E} \partial_{\bm{p}}  f_{eq}$, with $f_{eq}$ being the equilibrium Fermi-Dirac distribution in the band with the Fermi surface, and $\tau_{\rm{tr}}$ being the transport mean free time. Since both the transport current and the axial magnetization are determined by the same transport electric field, we can exclude it to obtain a direct relationship between the axial magnetization and the transport current: 
\begin{align}\label{eq:axialmagnetization}
\bm M_5\equiv \bm M_+-\bm M_{-} =\frac{\hbar}{2p_F}\bm j_{\rm{tr}}.
\end{align}

Using $\bm j_5=\bm \nabla\times \bm M_5$, and combining Eq.~\eqref{eq:axialmagnetization} with Eqs.~\eqref{eq:effectivefield} and~\eqref{eq:spinfromaxialcurrent}, we obtain the final expression for the current-induced effective Zeeman field in a centrosymmetric magnetic Weyl metal: 
\begin{align}\label{eq:finaleffectivefield}
    \bm B_{\rm{eff}}=\frac{\hbar J}{2e\epsilon_F M_s}\bm \nabla\times \bm j_{\rm{tr}},
\end{align}
where $\epsilon_F\equiv p_Fv$ is the Fermi energy counted from the energy of the Weyl nodes. 

Eq.~\eqref{eq:finaleffectivefield} for the effective Zeeman field is one of the central results of this work. Being determined by the curl of the transport current, this field vanishes in the bulk of an isotropic system, for which $\bm j_{\rm{tr}}=\sigma_D \bm E$, because of the Faraday's law for a static electric field, $\bm \nabla\times\bm E=0$. For an anisotropic model, in which the conductivity is a nontrivial tensor, this field can exist even in the bulk of the system if the electric field is nonuniform. But in any case the effective field is nonzero near a boundary of a sample, if there is a flow of current along the boundary. Another important feature of Eq.~\eqref{eq:finaleffectivefield} is that the magnitude of the effective field acting at the boundaries of a sample does not depend on the sample size, as long as spatial quantization is not important. 

We can gain further insight into the energy associated with the effective magnetic field, 
\begin{align}\label{eq:effectiveZeeman}
    E^Z_{\rm{eff}}=-\int_{\bm r}\bm B_{\rm{eff}}\cdot\bm M, 
\end{align}
if we perform an integration by parts over a volume bounded by a surface outside the sample, over which the magnetization vanishes. We then trivially obtain 
\begin{align}\label{eq:axialMfield}
    E^Z_{\rm{eff}}=-\int_{\bm r}\bm B_{5}\cdot\bm M_5, 
\end{align}
where the pseudomagnetic field $\bm B_5$ is given by Eq.~\eqref{eq:B5}, and the current-induced axial magnetization is given by Eq.~\eqref{eq:axialmagnetization}. Hence the energy that we obtained is nothing but the Zeeman energy of the two axial magnetizations in the corresponding axial magnetic field due to a magnetic texture. 

Before switching to applications, we would like to give another form of $E^Z_{\rm{eff}}$, appropriate for a sample with curl of the transport current confined to its surface: 
\begin{align}\label{eq:surfacefield}
    E^Z_{\rm{eff}}=-\frac{\hbar J}{2e\epsilon_F}\oint_S\,  \bm m\cdot \bm j_{\rm{tr}}\times\bm n. 
\end{align}
The surface integral in the last term, representing the effective Zeeman energy, runs over the entire sample surface, and $\bm n$ is the outer normal to the surface element $dS$. Expression~\eqref{eq:surfacefield} shows that the effective field~\eqref{eq:finaleffectivefield} is not unique in its form: an analogous contribution would come from the spin Hall effect, see the Discussion part at the end of this paper. Our point is that this field in Weyl metals is strong enough to control magnetic textures even without a spin Hall effect. Conversely, if the spin Hall effect is being studied in a magnetic Weyl material, it should be kept in mind that the current-induced axial magnetization can affect interpretation of experiments.

Finally, we would like to address the question of what limits the magnitude of the effective field. As is clear from Eq.~\eqref{eq:finaleffectivefield}, the maximum magnitude of the effective field is set by the maximum current one can drive through the sample. In a Weyl system, the maximum current in the linear regime is limited by the condition that the drift speed be smaller than the Fermi speed of Weyl electrons. In other words, the current is limited by $j_{\rm{max}}=en_{\rm W}v$, where $n_{w}$ is the total density of the Weyl electrons. Then from Eq.~\eqref{eq:finaleffectivefield} it follows that the maximum effective field scales as $B^{\rm{max}}_{\rm{\rm{eff}}}\propto\epsilon_F^2$, and saturates at $\epsilon_F\sim J$, where the Fermi surfaces near the two nodes go through a Lifshitz transition into a single trivial Fermi surface.

\textit{Magnetic vortex control in a Weyl nanodisk.---}
\begin{figure}
    \centering
    \includegraphics[width=2in]{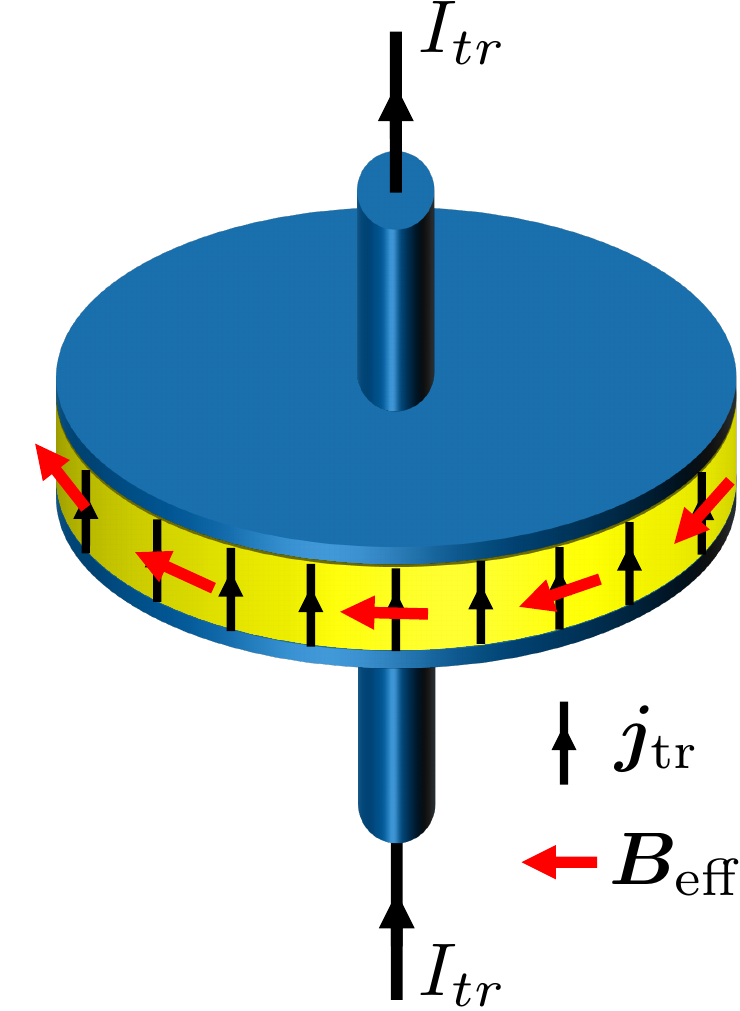}
    \caption{Schematic drawing of a magnetic nanodisk (thick yellow disk) with two leads (thin dark blue disks) that feed in transport current $I_{\rm{tr}}$. The upward black arrows show the direction of transport current inside the disk, and the thick red arrows winding around the disk boundary represent the effective magnetic field induced by the current.}
    \label{fig:Disk}
\end{figure}
We now show that the current-induced effective Zeeman field is also effective in the sense of magnetization control. We consider a thin metallic disk shown in Fig.\ref{fig:Disk}, in which a transport current is setup perpendicular to the plane of the disk. This current setup differs from the one considered for magnetic texture control in Ref.~\cite{kurebayashi2019theory}, where the current flow was in the plane of the disk. We assume that the transport current is reasonably uniform in the bulk of the disk, and is mostly perpendicular to the top and bottom surfaces of the disk. In this case the effective Zeeman field acts on the side surface of the disk, see Eq.~\eqref{eq:surfacefield}, and Fig.~\ref{fig:Disk}. As is seen from Eq.~\eqref{eq:surfacefield}, for $Jv>0$, the field obeys the \emph{left} hand rule, opposite to the Oersted field created by the current, because $e<0$. We will neglect the {\O}rsted field for the time being, but later will show that for disks of sizes measured in tens of nanometers the effect of the {\O}rsted field is small as compared to the effective field considered in this paper. 

Given the setup describe above, it is clear that the effective Zeeman field gives preference to certain chirality of magnetic vortices, and can switch between different chiralities for strong enough transport currents. Below we describe this process quantitatively.

We will assume that the magnetic energy of the disk, $E_M$, contains an exchange part associated with magnetization gradients, a dipolar part defined by the demagnetization field $\bm H_d$, and the effective Zeeman part, Eq.~\eqref{eq:effectiveZeeman}, in the presence of a current: 
\begin{align}\label{eq:magneticenergy}
    E_M=A\int_{\bm r}\nabla_a\bm m\nabla_a \bm m-\frac{\mu_0 M_s}2\int_{\bm r} \bm m\cdot \bm H_d+E^Z_{\rm{eff}}.
\end{align}
Below we will use the value of $A=10^{-11}\, \mathrm{J/m}$ for the exchange constant, $\mu_0M_s=1\,\mathrm{T}$ for the saturation magnetization, and $J/\epsilon_F\sim 10$ in the expression for the effective Zeeman energy, Eq.~\eqref{eq:surfacefield}. For these numbers the magnetic exchange length is $\ell_{\rm{ex}}=\sqrt{2A/\mu_0M_s^2}\approx 5\,\rm{nm}$.
We neglect the {\O}rsted field of the current, as its effect is small for the sizes of the disks considered, which we checked numerically.

\begin{figure}
    \centering
    \includegraphics[width=1\linewidth]{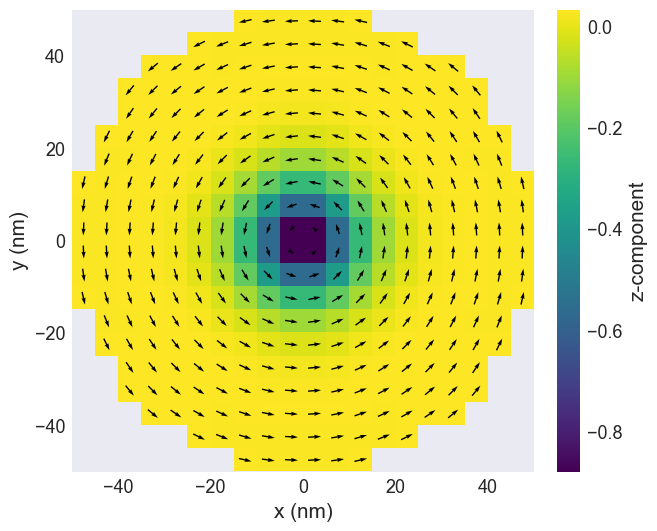}
    \caption{Magnetic vortex configuration in a disk of thickness $25\,\rm{nm}$ and radius $50\,\rm{nm}$. The polarity of the vortex is $-1$(topological charge $-\frac{1}{2}$), and its chirality is $+1$.}
    \label{fig:vortex}
\end{figure}

A disk of large enough radius contains a magnetic vortex~\cite{brown1963micromagnetics} of the in-plane magnetization, see Fig.~\ref{fig:vortex}. The vortex develops to minimize the dipolar energy at the expense of an increase in the exchange energy. To keep the exchange energy finite, a vortex must have a core with out-of-plane magnetization. Then a vortex is characterized with two discrete indices each taking values $\pm1$: the chirality of magnetization winding away from the core, and the polarization direction of the core. The four possible combinations of these indices are all degenerate for the energy~\eqref{eq:magneticenergy} in the absence of a transport current. 

It is worth noting that magnetic vortices of the described kind have definite positive winding for either sign of the chirality, in the sense that the azimuthal angle of the magnetization, $\phi$, winds in the positive direction with the azimuthal angle of the cylindrical coordinate system in real space, $\alpha$, the $z$-axis of which goes through the center of the disk, perpendicular to its plane:
\begin{align}
    \phi=\alpha+\pi/2+\eta.
\end{align}
In the equation for the azimuthal angle of the magnetization the quantity $\eta=0,\pi$ correspond to positive and negative chirality, respectively. An anti-vortex with negative winding would create magnetization pattern with nonzero radial component at the disk side surface, and thus would have high magnetostatic energy due to the magnetic charges on that surface. 

The fixed winding makes the topological index of the vortex, or its skyrmion charge, 
\begin{align}
    N(z)=\frac{1}{4\pi}\int dxdy\,\bm m(\bm r)\cdot(\partial_x\bm m(\bm r)\times\partial_y \bm m(\bm r)),
\end{align}
dependent on the vortex core polarization only.  Since the sample is three-dimensional, one can only define the topological charge for $z=\mathrm{const}$ plane, and the result is $z$-dependent. However, we checked numerically that even for a disk of diameter only twice as large as its thickness the skyrmion charge $N(z)$ as a function of $z$ does not deviate from the values of $\pm 1/2$ by more than $5\%$, so we are dealing with well-defined vortices.

We define the chirality as the volume integral over the interior of the sample, not including its boundary, of the $z$-component of the magnetization direction curl: 
\begin{align}\label{eq:chiralitydefinition}
    C=\frac{1}{2\pi r d}\int d^3 r\, \bm e_z\cdot\nabla\times \bm m.
\end{align}
This expression saturates at $\pm 1$ for a vortex with independent of the $z$-coordinate $\bm m$, which lies in the $xy$-plane near the sample boundary. In practice, for small and thin disks these conditions are satisfied in practice with high accuracy.

Without a current, the two vortex chiralities are degenerate in energy. It follows from the chirality definition~\eqref{eq:chiralitydefinition} and Eq.~\eqref{eq:axialMfield} for the effective Zeeman energy expressed via current-induced axial magnetization, as well as Eq.~\eqref{eq:axialmagnetization} for the axial magnetization itself, that for a transport current along the disk axis the effective Zeeman energy is proportional to the average disk chirality. Hence it makes one of the chirality states metastable. The {\O}rsted field of the current would have the same qualitative effect, but for disk diameters around hundred nanometers the effect of the {\O}rsted field is small.  For large enough current the metastability is removed, and the effective field induces deterministic switching into the low-energy state. 

A rigorous analytic way to determine the critical switching current would be to perform the linear stability analysis of the vortex excitation modes~\cite{guslienko2005modes}. To this end one calculates the eigenmodes of small magnetization oscillations, and finds the value of the effective boundary field that drives the lowest frequency to zero. The zero-frequency mode becomes the nucleation one~\cite{brown1963micromagnetics}, along which the chirality reversal proceeds. This analysis is very involved due to complicated patterns of demagnetizing fields.  We thus proceed with a numerical analysis of the switching current.

\begin{figure}
    \centering
    \includegraphics[width=1\linewidth]{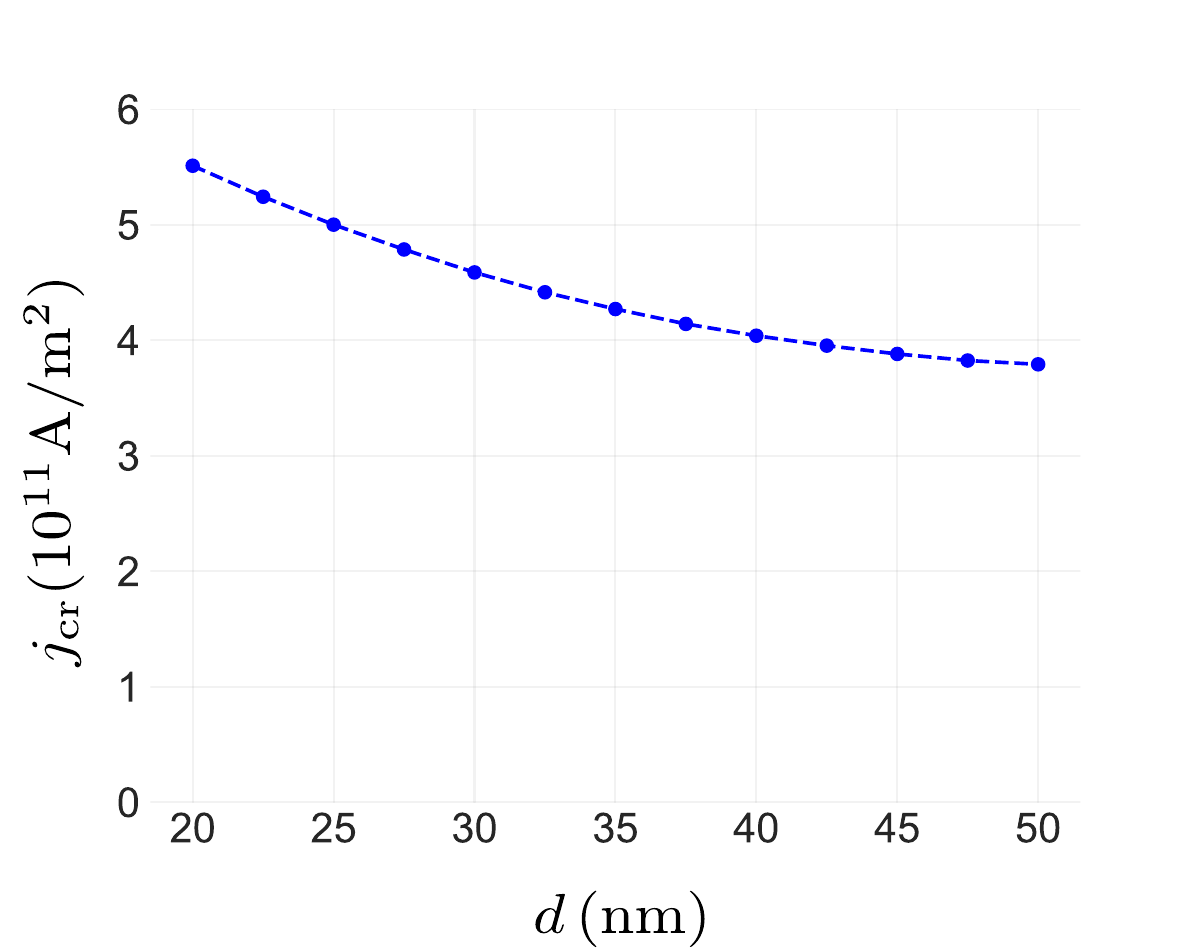}
    \caption{Critical switching current, $j_{\rm{cr}}$, for a disk of 50nm radius as a function of its thickness, $d$.}
    \label{fig:critcurrent}
\end{figure}

To obtain the critical current for the chirality switching, we simulated the system dynamics with slowly varying values of the transport current to determine the value at which the chirality switches. We did not attempt to simulate realistic temporal dynamics for some current pulses. 
\begin{figure}
    \centering
    \includegraphics[width=3.5in]{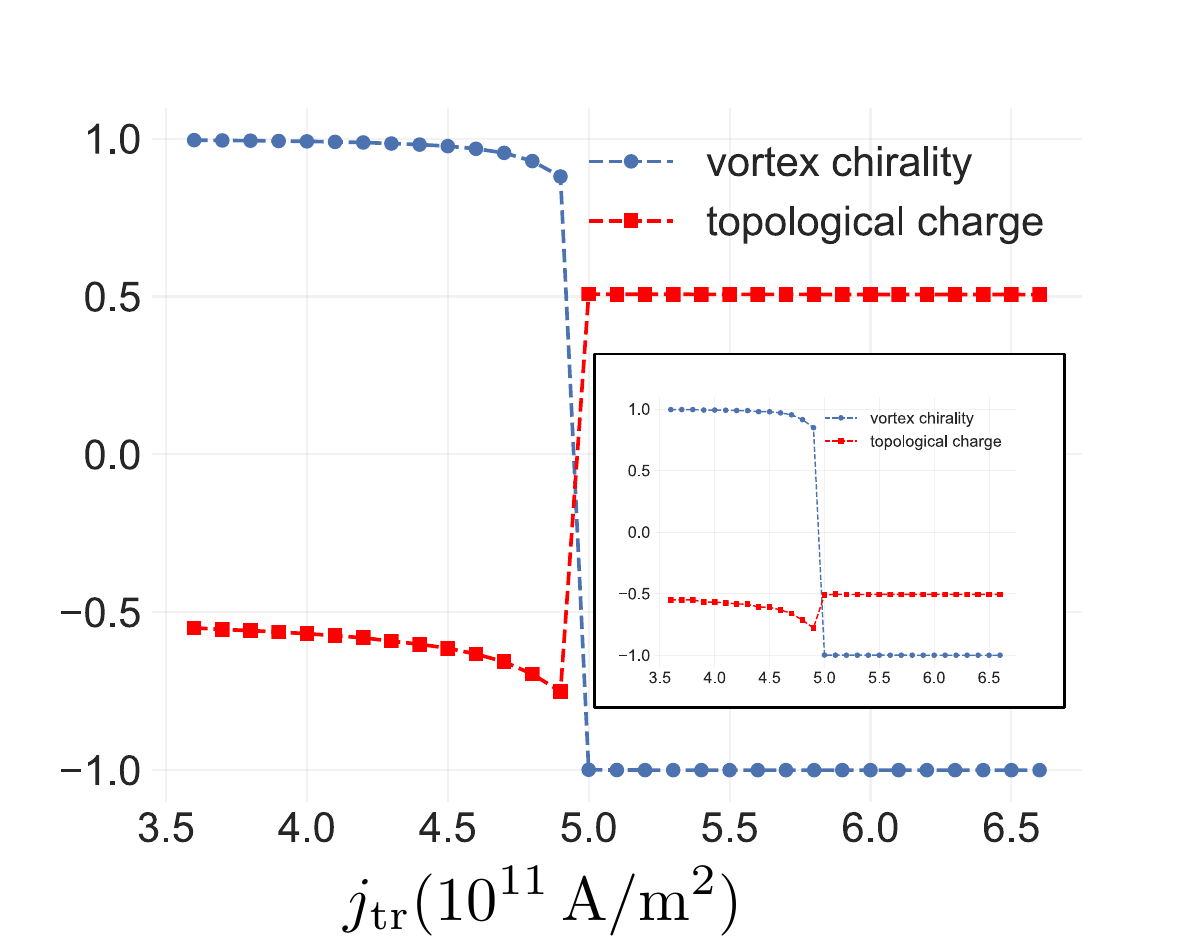}
    \caption{(Color online) Chirality of a magnetic vortex and its topological charge as a function of transport current density, $j_{\rm{tr}}$, as it is gradually increased. Both the main panel and the inset correspond to numerical simulations of the Landau-Lifshitz-Gilbert equation for a disk sample with the thickness of $25\,\mathrm{nm}$, and the radius of $50\,\mathrm{nm}$. In the main panel the Gilbert damping constant is $\alpha=0.01$, while in the inset $\alpha=0.001$. The small value of $\alpha$ did not lead to polarization reversal, while the critical current for chirality reversal remained the same.}
    \label{fig:chirality}
\end{figure}
The results of the simulation for a disk of radius $50\,\mathrm{nm}$ and variable thickness are shown in Fig.~\ref{fig:critcurrent}. We used  We obtained critical currents of the order of $5\times 10^{11}\,\mathrm{A/m^2}$, which are feasible from the practical point of view. Equating the value of the critical current to the maximum achievable current in the linear regime, $j_{\rm{max}}=e n_{\rm W}v$, and using $v=10^5\,\mathrm{m/s}$, we see that the required carrier density is $n_{\rm W}\sim{10^{19}}\,\mathrm{cm}^{-3}$. Hole doping of ${10^{20}}\,\mathrm{cm}^{-3}$ in EuCd$_2$As$_2$ was reported in Ref.~\cite{roychowdhury2023eucd2as2}.

We also noticed empirically that for relatively large values of the Gilbert damping constant the polarization of the core switched together with the chirality in small disks. The typical graphs of the chirality and the Skyrmion number as functions of the applied static current are shown in Fig.~\ref{fig:chirality}. Note that with decreasing value of $\alpha$ the polarization fails to switch, while the critical current does not change. This shows that the polarization switching is a dynamic effect, which is sensitive to the speed of the chirality reversal, while the chirality itself switches when it loses metastability, regardless of how fast the subsequent dynamics is.

Finally, we note that for pure {\O}rsted field of the current, neglecting the effective boundary field, the critical switching current for the geometry considered here is roughly $5\times 10^{12}\,\mathrm{A/m^2}$, and order of magnitude larger than for the boundary field. This is consisted with the findings of Ref.~\cite{aldulaimi2023oersted}, and shows that neglecting this field was justified for our purposes. Of course, for large enough disks the {\O}rsted field will eventually dominate the switching.

\textit{Discussion.---} The central result of this work is the observation that a transport current flowing in a centrosymmetric magnetic Weyl metal induces axial magnetization. The induced axial magnetization currents correspond to the spin polarization of itinerant electrons. This spin polarization can be used to control the chirality of a vortex in the magnetization of localized electrons via an effective Zeeman field, Eq.~\eqref{eq:finaleffectivefield}. 

It is interesting to compare this mechanism with proposals to generate axial currents, and hence itinerant spin polarization, in current-carrying Weyl metals in the existing literature. In Ref.~\cite{kurebayashi2019theory} it was shown that axial Hall effect, driven by the pseudomagnetic field $\bm B_5$, drives an axial Hall current $\bm j_5\propto \bm B_5\times \bm j_{\rm{tr}}$. Later in Ref.~\cite{hannukainen2021electric} the axial version of the chiral magnetic effect was used in conjunction with the chiral anomaly to generate $\bm j_5\propto \bm B_5(\bm B\cdot \bm j_{\rm{tr}})$, where $\bm B$ is the external (but which can be the field of the magnetization itself) magnetic field driving the chiral anomaly.  In contrast, in this work the axial current takes the form of $\bm j_5\propto \nabla\times \bm j_{\rm{tr}}$. This axial current, unlike those from Refs.~\cite{kurebayashi2019theory,hannukainen2021electric}, is not proportional to $B_5$, Eq.~\eqref{eq:B5}. This makes it at least one or maybe two orders of magnitude smaller than the other two axial currents, since $B_5$ is large due to the large value of the exchange constant and small exchange length that determines the size of magnetic textures in ferromagnets. However, its independence from $\bm B_5$ is also its strength from the symmetry point of view: the axial current considered here is even in the localized magnetization, and hence can distinguish chiralities of a magnetic vortex. As we demonstrated, the magnitude of the effect is sufficient to drive chirality reversals in nanosized samples more efficiently than with the {\O}rsted field of the current. 

It also interesting to compare the boundary spin polarization associated with the axial magnetization current to the one that would have been induced by an isotropic spin Hall effect, if it existed in the sample. In that case the spin polarization current is given by $j^a_b=\theta\epsilon_{abc}j_{\rm{tr},c}/e$, where $j^a_b$ is the current of $a$th component of spin polarization in the $b$th spatial direction, and $\theta$ is the spin-Hall angle. Then for electric current flowing in the $z$-direction along boundary perpendicular to the $x$-direction there is a spin accumulation of surface density of the $y$th component of spin polarization of magnitude $\sim\tau_{\rm{sf}}\theta j_{\rm{tr}}/e$. This result needs to be compared to the spin accumulation given by the current-induced axial magnetization current, given by $\sim \hbar j_{\rm{tr}}/e\epsilon_F$. For instance, for Pt $\tau_{\rm{sf}}\sim 10^{-14}\,\mathrm{s}$~\cite{urazhdin2018spinrelaxation}, and $\theta\sim 10^{-1}$~\cite{hoffmann2013spindiffusion}, which yields $\tau_{\rm{sf}}\theta\sim 10^{-15}\,\mathrm{s}$. For the mechanism described in this work and the typical $\epsilon_F\sim 50\mathrm{meV}$ we obtain $\hbar/\epsilon_F\sim 10^{-14}\,\mathrm{s}$, obviously implying a much larger boundary spin polarization. This order of magnitude larger boundary spin polarization may even be utilized in the spintronics applications.

The work of JGY and DAP was supported by the National Science
Foundation under Grant No. DMR-2138008 . The work of YT was supported by the U.S. Department of Energy, Office of Basic Energy SCiences under Award No. DE-SC0012190.

\bibliography{references}
\bibliographystyle{apsrev}
\end{document}